\documentclass[fleqn,twoside]{article}
\usepackage[headings]{espcrc2}
\usepackage{graphicx}

\readRCS
$Id: espcrc2.tex,v 1.2 2004/02/24 11:22:11 spepping Exp $
\usepackage{graphicx, balance}
\usepackage[figuresright]{rotating}

\newcommand{\AmS}{{\protect\the\textfont2
  A\kern-.1667em\lower.5ex\hbox{M}\kern-.125emS}}

\title{\textbf{Improved Bully Election Algorithm for Distributed Systems}}

\author{P  Beaulah  Soundarabai\address[DCSE]{Department of Computer Science , Christ University, Bangalore 560 029 India,\\~ Contact: beaulah.s@christuniversity.in\\},
Ritesh Sahai\addressmark,
Thriveni J\address{University Visvesvaraya College of Engineering, Bangalore University, Bangalore.\\},
K  R  Venugopal\addressmark,
L  M  Patnaik\address{Honorary Professor, Indian Institute of Science, Bangalore.}}

\runtitle{Improved Bully Election Algorithm for Distributed Systems}
\runauthor{P  Beaulah  Soundarabai, $et~ al.,$}

\begin{document}
\begin{abstract}

Electing a leader is a classical problem in distributed computing system. Synchronization between processes often requires one process acting as a coordinator. If an elected leader node fails, the other nodes of the system need to elect another leader without much wasting of time. The bully algorithm is a classical approach for electing a leader in a synchronous distributed computing system, which is used to determine the process with highest priority number as the coordinator. In this paper, we have discussed the limitations of Bully algorithm and proposed a simple and efficient method for the Bully algorithm which reduces the number of messages during the election. Our analytical simulation shows that, our proposed algorithm is more efficient than the Bully algorithm with fewer messages passing and fewer stages.  
\end{abstract}

\maketitle

\section{INTRODUCTION}
Distributed computing is a decentralized and parallel computing, using two or more computers communicating over a network to accomplish a common task. Centralized control in distributed systems helps to achieve some specific goals such as mutual exclusion, synchronization, load balancing, and time scheduling. This type of distributed system often requires a unique node to play the role of leader or coordinator of the other nodes to take care of synchronization. As node crash failure is very common in distributed systems. Failure of a leader node requires special attention and needs extra tasks to elect another one to act as leader.
\vskip 2mm
The collaborating processes are often identical. One of the central problems is election of a leader. Given a network of processes, exactly one process should take the decision that it is the leader. It is usually required that all non-leader processes are informed or involved in the process of the leader election. A leader election algorithm is one of the basic activities of distributed systems, as it acts as a basis for more complex and high level algorithms and applications. An important challenge in distributed systems is the adoption of suitable and efficient algorithms for coordinator election. The main role of an elected coordinator is to manage the use of a shared resource in an optimal manner which in turn maintains the coherency of the system even during partial failures.
\subsection{Motivation}
The main drawback of Bully algorithm is more number of message passing. As it is mentioned before the message passing has order O($n^2$) that increases traffic in network. It also has five stages to decide the next leader which would waste a lots of time for the processes to resume their normal execution. Bully algorithm is a safe way for election; however its traffic is relatively high.
\subsection{Contribution}
In this paper, we have proposed a modified Bully algorithm which preserves all the advantages of the existing algorithm and at the same time eliminates the limitations of it by reducing the number of messages and the the number of stages to elect the next leader.
\subsection{Organization}
The remainder of this paper is organized as follows: Section 2 reviews the related work, Section 3 describes the Problem Definition and Methodology of Bully algorithm, Modified Bully algorithm is given in Section 4, Section 5 details the simulation and the comparison of the two algorithms, Conclusions are presented in Section 6.
\section{\uppercase{Literature Survey}}
Effat Parvar M R \cite{1} described  novel approaches towards improving the Bully and Ring algorithms and also proposed the heap tree mechanism for electing the coordinator. The higher efficiency and better performance with respect to the existing algorithms was also validated through simulation.
\vskip 2mm
Sandipan Basu \cite{2} has discussed the limitations of bully algorithm and proposed a modified algorithm. In the original bully algorithm, when the leader process is crashed, immediately the new leader is elected. But, if the old leader process comes back, it once again initiates the election. The author suggests that there need not be another election, instead, the old leader process can accept the new leader process by sending the new request of who the leader is?, to its neighbor. In the next round of election, it can try becoming the leader.
\vskip 2mm
Muhammad Mahbubur Rahman $et~al.,$ \cite{3} have also proposed a modified bully election algorithm. In their paper, they say that the bully algorithm has O($n^2$) messages which increases the network traffic. In the worst case, n number of elections can occur in the system which again in turn will yield in a heavy network traffic. They have proposed the same algorithm but with Failure Detector, Helper processes to have unique election with the Election Commission.
\vskip 2mm
Chang-Young Kim $et~al.,$ \cite{4} have proposed the election protocol for reconfigurable distributed systems which again was based on bully election algorithm. The actual election is run by the base stations making the protocol, energy efficient. The protocol is also independent from the overall number of mobile hosts and the data structures required by the algorithm are managed at the base station, making the protocol scalable as well.
\vskip 2mm
M S Kordafshari $et~al.,$ \cite{5} have done a survey of  synchronous bully algorithm and modified it with an optimal message algorithm and also discussed the limitations of these algorithms. The authors have tried to reduce the number of elections happening in the classical bully algorithm. The proposed algorithm has only one election at any point of time, which brings down the number of messages being exchanged drastically. Sepehri M $et~al.,$ \cite{6} have dealt with the distributed leader election algorithm for a set of processes connected by a tree network. The authors have proposed a linear time algorithm using heap structure using reheap up and reheap down algorithms. They also have analysed the algorithm and reached a logarithmic number of message complexity.
\vskip 2mm
Sung Hoon Park \cite{7} has proposed Failure Detector which has an overhead module with a specialised function that detects crash and recovery of a node in a system. This report can be given to any process at request. The author modified the bully algorithm using the failure detector. The performance of the system goes down because of the overhead of Failure Detector. Failure Detector is the centralized component, which leads to the traditional problems of single point of failure and bottleneck scenario of single queue access the resource.
\vskip 2mm
Zargarnataj \cite{8} has developed an algorithm which is based on Bully’s election algorithm with an additional feature of an assistant to the new leader. If the present leader node crashes, the assistant leader would become the new leader without any overhead of election. Whenever any process realises the absence of the leader, it immediately sends a message to the assistant leader to alert it. When the assistant leader receives any such message, it confirms the unavailability of the leader by timeout message and if it is true, it broadcasts the leader message to all the processes. But the limitation of this algorithm is that if the assistant leader is also not available then there is a lot of messages sent and time delays to invoke the election algorithm once again.
\section{\uppercase{Bully Algorithm}}
Distributed systems require some special capabilities of a good and efficient leader election algorithm, such as leader longevity, low communication overhead, low complexity in terms of time and messages, and providing uniqueness to the elected leader. Several algorithms have been proposed to deal with the leader node failure problem, and the Bully Algorithm is the classical one amongst them for electing a leader node in synchronous systems, although this algorithm demands a large number of messages between the nodes.
\vskip 2mm
Bully algorithm was first presented by Garcia Molina in 1982. The Bully algorithm in distributed computing system is used for dynamically electing a leader by using the process ID number. The process with the highest process ID number is elected as the leader process.
\vskip 2mm
1. Assumptions
\renewcommand{\labelenumi}{(\roman{enumi})}
\begin{enumerate}
\item Each process has a unique and not null number to distinguish them and each process knows other process’ number.
\item Processes don’t know which ones are currently up and down.
\item The entire system is synchronous. Timeouts are used for deciding process failure.
\item Processes can crash even during execution of algorithm.
\item Message delivery between processes is reliable and time bound.
\end{enumerate}

2. Aim:\\
The aim of election Algorithm execution is selecting one process as leader (Coordinator) that all processes agree with it. In other words, electing a process with the highest priority or highest ID number as a leader or coordinator.\\
Suppose that the process P finds out the coordinator crashed, P immediately holds an election. This algorithm has the following steps:\\
a) Step1:\\
When a process, P, notices that the coordinator crashed, it initiates an election algorithm.
\renewcommand{\labelenumi}{(\roman{enumi})}
\begin{enumerate}
\item P sends an ELECTION message to all processes with higher numbers respect to it.
\item If no one responses within the time limit, P wins the election and becomes a coordinator.
\end{enumerate}
b) Step2:\\
When a process receives an ELECTION message from one of the processes with a lower number response to it:
\renewcommand{\labelenumi}{(\roman{enumi})}
\begin{enumerate}
\item The receiver sends an OK message back to the sender to indicate that it is alive and will take over.
\item The receiver holds an election, unless it is already holding one.
\item Finally, all processes give up except one that is the new coordinator.
\item The new coordinator announces its victory by sending a message to all processes telling them, it is the new coordinator.
\end{enumerate}
c) Step3:\\
Immediately after the process with higher number compare to coordinator is up, bully algorithm is run.\\
\begin{figure*}
\parbox{18cm}{
\centering
\includegraphics[width=24pc, height=22pc]{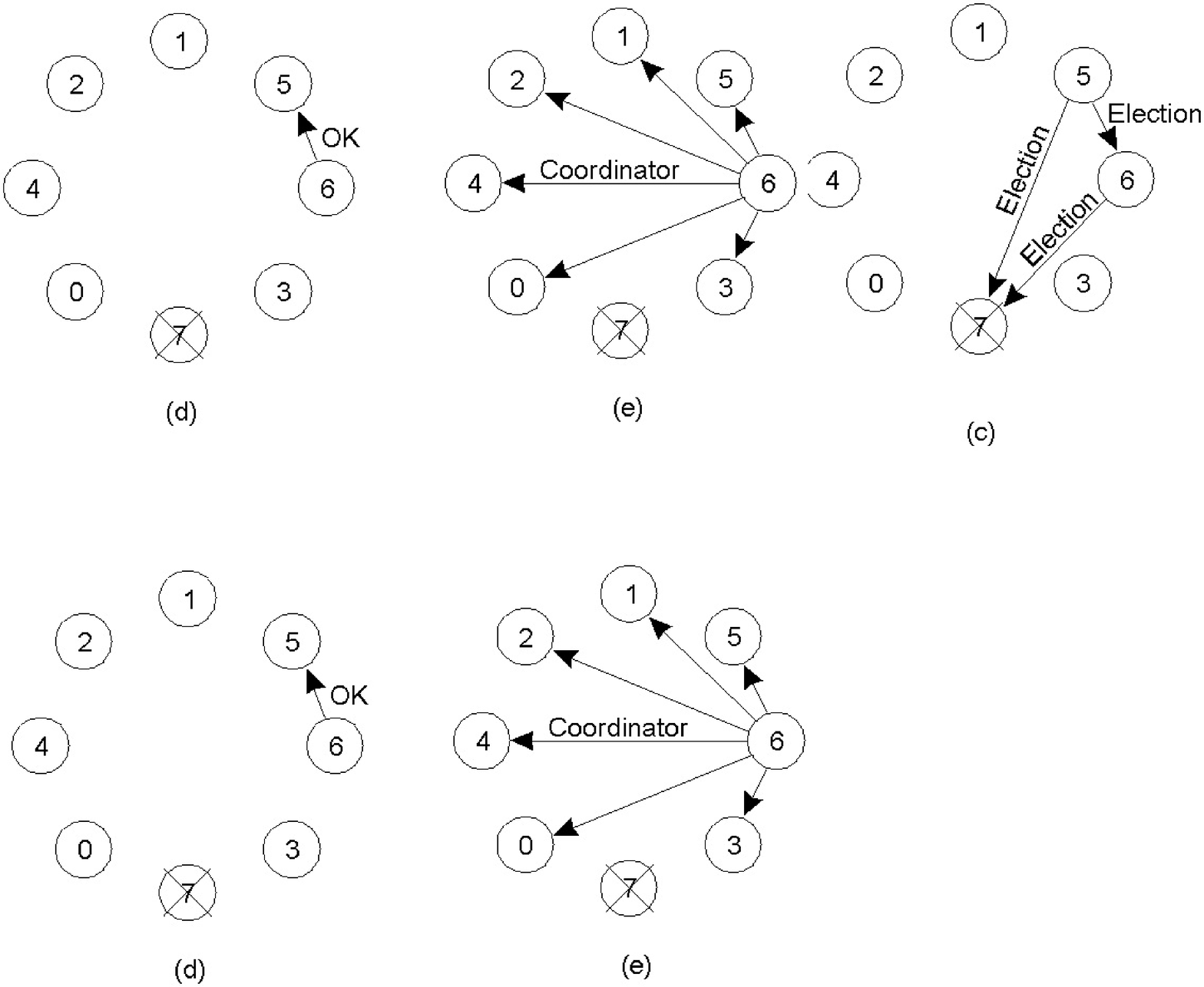}
\caption{Bully Algorithm.}}
\end{figure*}

Figure 1 explains the steps involved in the Bully election algorithm.\\
\renewcommand{\labelenumi}{(\roman{enumi})}
\begin{enumerate}
\item Process 4 holds an election
\item Process 5 and 6 respond, telling 4 to stop
\item Now 5 and 6 each hold an election individually leading to two simultaneous elections.
\item Process 6 tells 5 to stop by sending OK to it.
\item Process 6 wins the election because no higher processes responded to it and it informs all the processes that it is the Coordinator from now on.
\end{enumerate}

\subsection{Advantages and limitations.}
The advantages of Bully algorithm are that this algorithm is a distributed method with simple implementation \cite{9}\cite{10}\cite{11}. This method requires at most five stages, and the probability of detecting a crashed process during the execution of algorithm is lowered in contrast to other algorithms. Therefore other algorithms impose heavy traffic in the network in contrast to Bully algorithm \cite{12}. Another advantage of this algorithm is that only the processes with higher priority number respect to the priority number of process that detects the crash coordinator will be involved in election, not all process are involved. However the two major limitations of Bully algorithm are the number of stages to decide the new leader and the huge number of messages exchanged due to the broadcasting of election and OK messages \cite{13}.
\section{\uppercase{Modified Bully Algorithm}}
Generally, in fault-tolerant distributed systems the leader node has to perform some specific controlling tasks and this node is well known to the other nodes. This node does not necessarily possess any extra processing feature to become elected, but having the highest process-ID. Election algorithms need a special mechanism to elect the leader. After crash failure of the leader node, it is urgently needed to reorganize the existing active nodes to call for an election and to elect a leader in order to continue the operation of the entire system.
\subsection{Modified Bully Algorithm Details}
1. Assumption\\
      Besides having all the assumptions of the existing algorithm, we assume
\renewcommand{\labelenumi}{(\roman{enumi})}
\begin{enumerate}
\item All processes hold an election flag, if this flag is true election cannot be initiated by any process.
\item All processes have a variable to store coordinator information.
\end{enumerate}
a) Step1\\
Initially all election flag are set to false. When a process, P, notices that the coordinator crashed, it initiates an election algorithm.
\renewcommand{\labelenumi}{(\roman{enumi})}
\begin{enumerate}
\item P sends an ELECTION message to all processes.
\item  All processes set their election flag to true, so that none of the process can start \item Coordinator variable reset to zero.
\item If no one responses, P wins the election and becomes a coordinator.
\end{enumerate}
b) Step2\\
When a process receives an ELECTION message from one of the processes with lower numbered response to it:
\renewcommand{\labelenumi}{(\roman{enumi})}
\begin{enumerate}
\item The receiver sends an OK message back to the sender to indicate that it is alive and will take over.
\item The sender P extracts process ID of receiver and store it in coordinator variable. Only IDs greater than the stored ID can override the coordinator ID variable value.
\item Finally, all processes responded and higher process ID among them is stored in coordinator variable.
\item The sender P collects coordinator ID from variable and informed him (coordinator process ID) that he is coordinator.
\item The elected coordinator process cross check with his higher processes, if any higher process is alive he will take over, else currently elected process will be coordinator.
\item The new coordinator announces its victory by sending a message to all processes telling them, it is the new coordinator.
\item All processes set the coordinator ID in coordinator variable and reset election flag to false.
\end{enumerate}
c) Step3\\
Immediately after the process with higher number compare to coordinator is up, bully algorithm is run.
\begin{figure*}[ht!]
\parbox{18cm}{
\centering
\includegraphics[width=24pc, height=20pc]{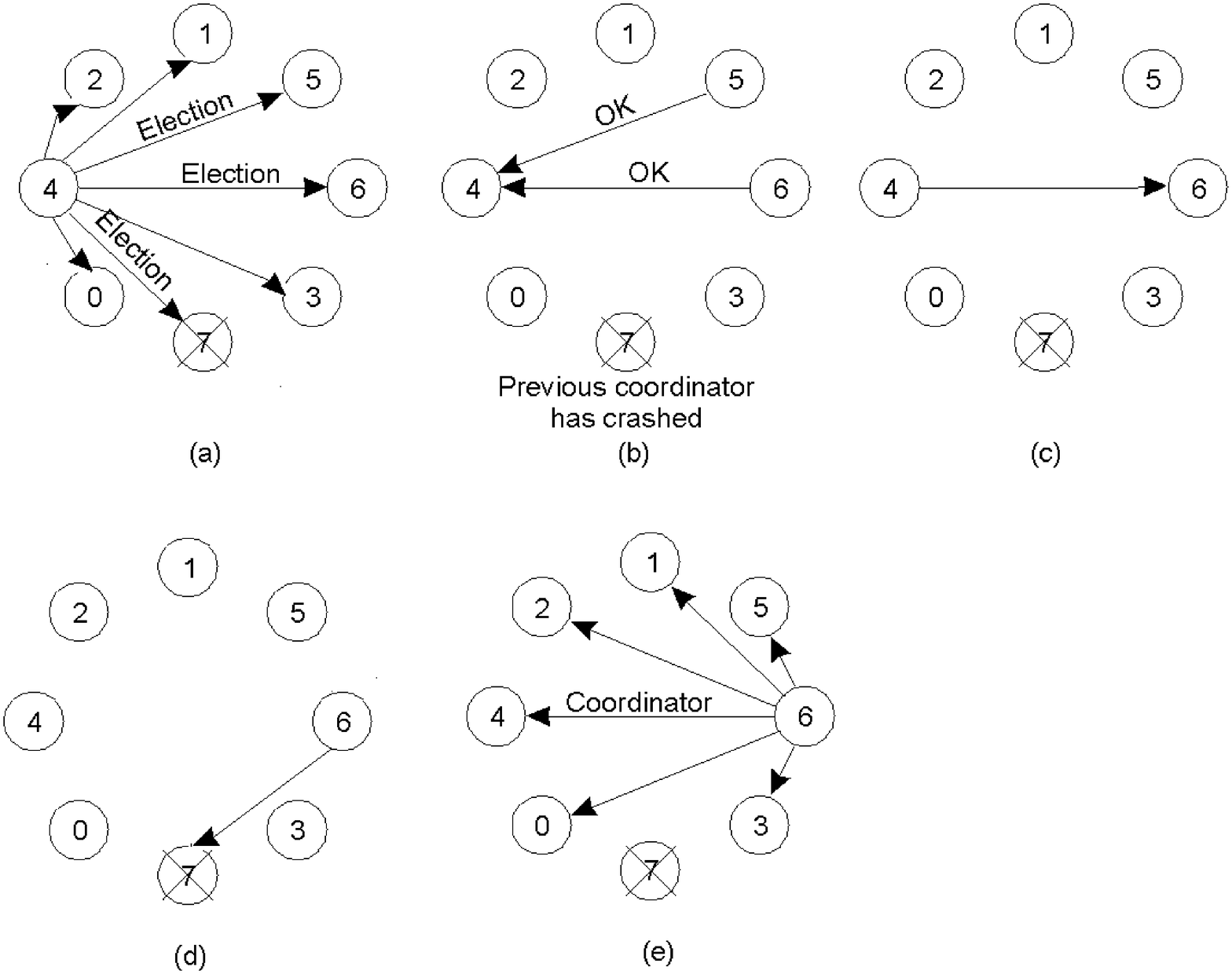}
\caption{Modified Bully Algorithm.}}
\end{figure*}

\begin{figure*}
\centering
\includegraphics[width=18pc, height=16pc]{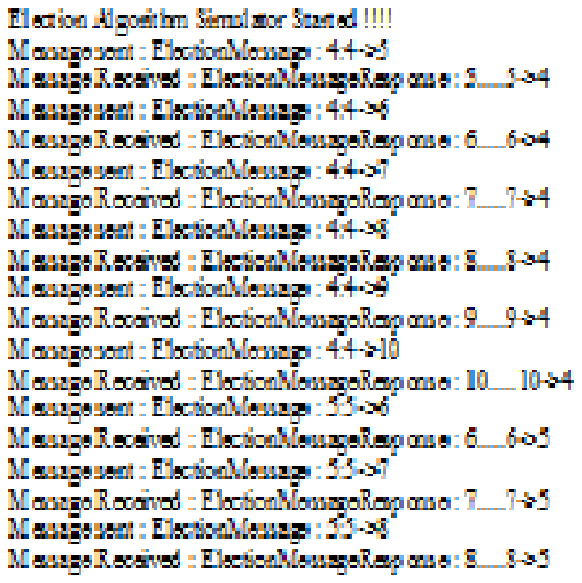}
\caption{Simulation Log1}
\centering
\includegraphics[width=34pc, height=26pc]{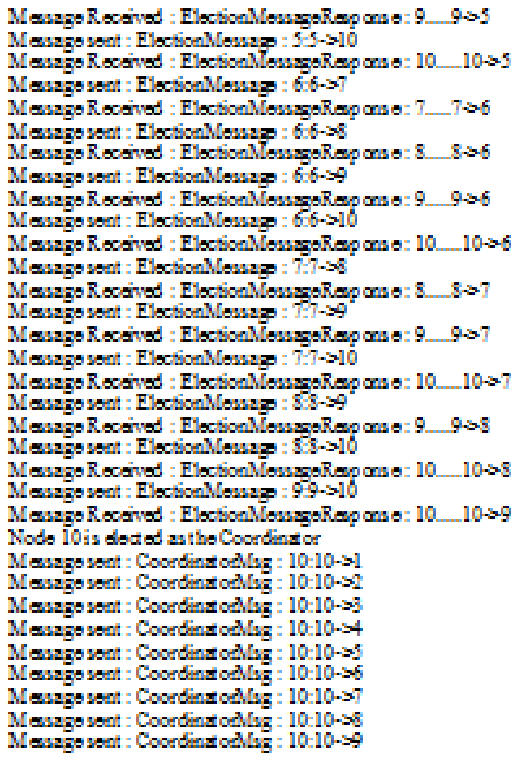}
\caption{Simulation Log2}
\end{figure*}
\begin{figure*}
\centering
\includegraphics[width=22pc, height=18pc]{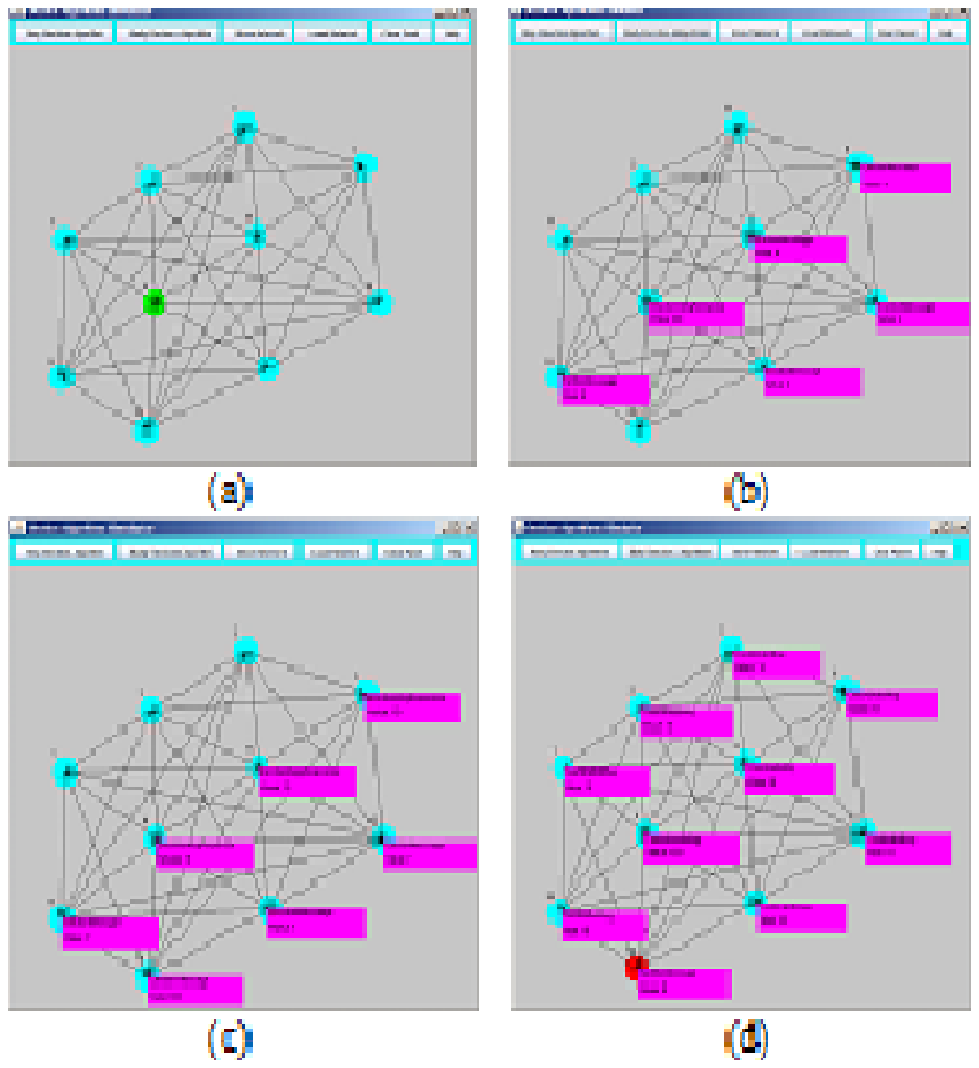}
\caption{ Bully Algorithm Simulation}
\end{figure*}

\vskip 2mm
Figure 2 shows the steps involved in Modified Bully Election Algorithm.
\renewcommand{\labelenumi}{(\roman{enumi})}
\begin{enumerate}
\item Process 4 holds an election
\item Process 5 and 6 respond, informing 4 about their presence in the system by OK message.
\item Processes 4  informs 6 to become coordinator.
\item Process 6 checks with process 7 if it is come back.
\item Since no reply from process 7, process 6 wins and broadcasts the Coordinator Message to all the processes.
\end{enumerate}
\subsection{Advantages}
Modified Bully algorithm is having all advantages of Bully algorithm. The additional advantages of modified Bully algorithm are that this algorithm is a very simple, having fail-safe mechanism, no parallel election, and reduced number of messages.
\subsection{Limitations}
How long the election initiator should wait to get response from all higher processes. If we keep a timer then the limitation could be the timeout value. Higher timeout will raise performance issue and lower timeout may miss responses from higher processes due to busy network traffic. However failsafe mechanism will be very helpful in this case.
\section{\uppercase{Simulation and Comparision}}
\subsection{Election Algorithm Simulator}
We used a GUI based simulator for simulating election algorithm. This simulator was capable of creating process node, creating distributed process network, message passing and electing the coordinator. The simulator is enhanced with GUI capability, allowing users to save and load the distributed network, display messages, selecting start node (green) and recognizing coordinator node by changing the color (red).
\subsection{Bully Algorithm Simulation}
\begin{enumerate}
\item Simulation Setup:\\
Started with 10 process nodes participating in election. Node 4 started election.
\item Bully Simulation Logs
\end{enumerate}
Figure 3 and Figure 4 shows the the Simulation logs of Bully Algorithm.
Figure 5 represents the simulation of Bully algorithm in which Figure 5(a) and Figure 5(b) shows that, process ID 4 identifies the absence of the leader and initiates the election  by sending the election message to its higher ups namely to processes 5, 6,\dots, 10. All these processes in turn start their own election and concludes the election by the coordinator message where process ID 10 is the new leader. These activities are depicted in Figure 5(c) and Figure 5(d).
\subsection{Modified Bully Algorithm Simulation}
\begin{enumerate}
\item Simulation Setup:\\
Started with 10 process nodes participating in election. Node 4 started election.
\item Modified Bully Simulation Logs
\end{enumerate}
\begin{figure*}
\centering
\includegraphics[width=28pc, height=22pc]{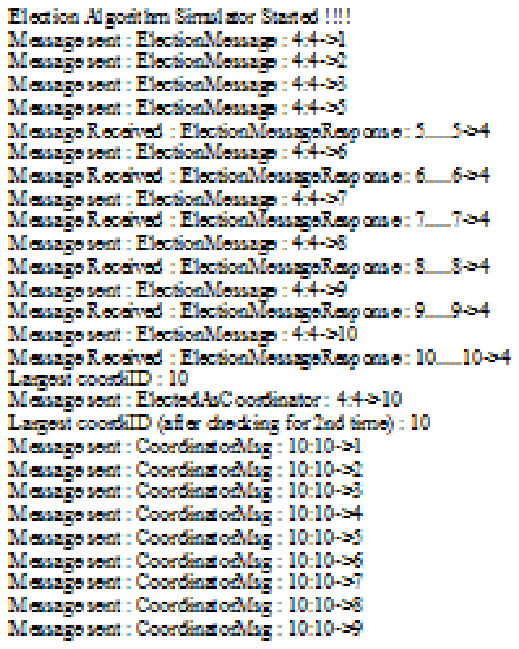}
\caption{ Modified Bully Algorithm Simulation Logs}\qquad
\centering
\includegraphics[width=25pc, height=18pc]{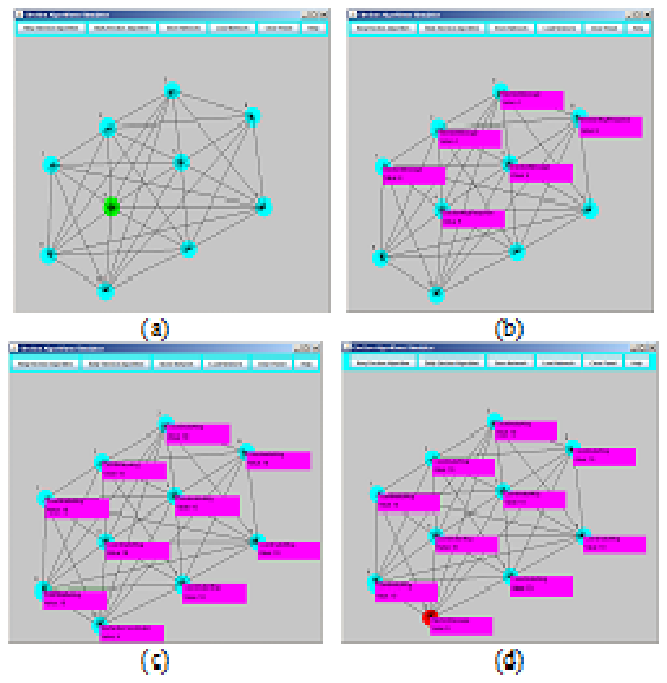}
\caption{ Modified Bully Algorithm Simulation}
\end{figure*}
The Figure 6 shows the Simulation logs of the Modified Bully Algorithm.The simulation of Bully algorithm is represented Figure 7, where Figure 7(a) and Figure 7(b) show that, process ID 4 identifies the absence of the leader and initiates the election  by sending the election message to its higher ups namely to processes 5, 6, \dots, 10. Unlike the Bully algorithm, all these processes reply to the initiator process 4 instead of starting their own election.  In Figure 7c) and Figure 7(d) show that process 4 decides the new coordinator (which is 10 in our simulation) and informs process 10 to take over and the election gets concluded by  the broadcast of coordinator message where process ID 10 is the new leader.
\begin{table}
\begin{center}
\caption{Message Comparison of Bully and Modified Bully Algorithms}
\small\addtolength{\tabcolsep}{-3pt}
\scalebox{0.8}{
\begin{tabular}{|c|c|c|}
 \hline
  \ & \bf Messages & \\
\hline
  \bf Processes & \bf Bully Algorithm & \bf Modified   \\
 & \bf  & \bf  Bully Algorithm  \\
  \hline
 5 & 24 & 13 \\
   \hline
  10 & 99 & 28 \\
  \hline
 15 & 224 & 43 \\
  \hline
  20 & 399 & 58\\
  \hline
  25 & 624 & 73 \\
  \hline
\end{tabular}}
\end{center}
\end{table}
\subsection{Message Comparison}
Table 1 shows the comparison for both algorithms. In this table we represented the message growth following by corresponding number of processes in the distributed network. Table 1 shows that numbers of messages are increasing drastically in the Bully algorithm compare to the modified Bully algorithm.
\vskip 2mm
Figure 8 shows a comparison graph where both Bully and modified Bully are highlighted in different colors. Graph presents the comparison where number of nodes represented by horizontal axis and number of messages represented by vertical axis. Graph shows that Bully is having curve shape that describe O(n$^2$) and modified Bully algorithm is having linear growth described by a straight line or O($n$).

\subsection{Simulation Result}
Our simulation result shows that modified election algorithm is more efficient as it reduces the number of messages, also avoided any parallel election process. The comparative results are well explained by simulation logs, comparison graph and table.
\begin{figure*}[ht!]
\centering
\includegraphics[width=22pc, height=16pc]{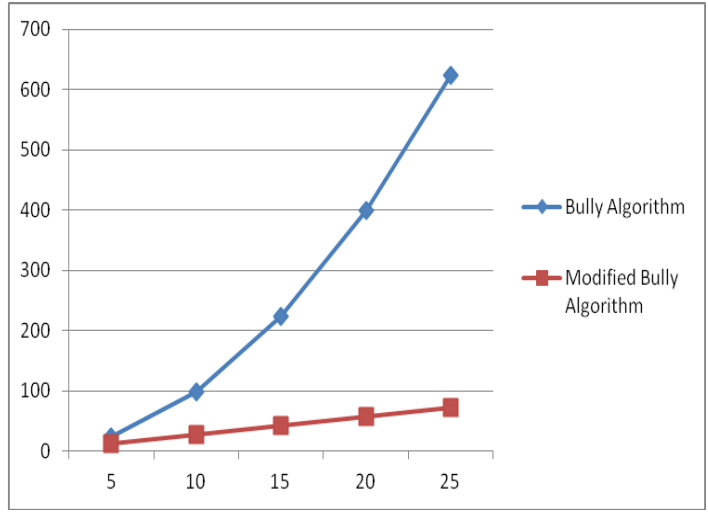}
\caption{ Number of Messages used During the Election}
\end{figure*}

\subsection{Analytical Comparision}
If only one process detects crashed coordinator\\
$N$: The number of processes\\
$P$: The priority number of processes that find out the crashed coordinator\\
$T_m$: The number of messages passing between processes when the Pth member detects the crashed Coordinator.\\
In Bully Modified algorithm the number of massages passing between processes for performing election is obtained from the following formula:
\begin{equation}
T_m= {2 * (N - P) + N }
\end{equation}

Which has Order O($n$). In the worst case that is $P$ = 1 (process with lowest priority number finds out crashed coordinator):
\begin{equation}
T_1 = 2 * (N - 1) + 1 = 3N-1
\end{equation}
Whereas the number of massage passing between processes in the Bully algorithm for performing election is obtained from the following formula:
\begin{equation}
T_m = (N - P + 1) (N - P) + N - 1
\end{equation}
In the worst case that is $P$= 1 (process with lowest priority number detects crashed coordinator):
\begin{equation}
T_1 = N^2 - 1
\end{equation}
Which has Order O($n^2$). Number of messages in proposed Bully algorithm will be equal to 3$n$ -1 that obviously means this modified algorithm is better than bully algorithm.
Now assume that the set of processes in $S$= \{$P_1$, $P_2$, $P_3$, ..., $P_n$\} from processes find out the crashed coordinator concurrently ($P_1$ is the lowest process).

In Bully algorithm, considering worst case and assuming lowest process start election, then:
\begin{enumerate}
\item Total number of election message sent to set ($S$) of $n$ processes (\{$P_1$, $P_2$, $P_3$, … , $P_n$\}) are ($n$ - 1).
\item Total response message received by $P_1$ is ($n$ - 1).
\item Now $P_2$ will send election message to $n$ - 2 processes.
\item Total response message received by $P_2$ is ($n$ - 2).
\item Similarly for $P_3$, $P_4$ and $P_n$.
\item Finally $P_n$ informing to every process by sending coordinator message is again ($n$ - 1) message.

The number of message passing between processes for  performing election is obtained from the following formula:

$T_m$ = ($n$ - 1) + ($n$ - 2) + ($n$ - 3) + \dots + ($n$ - $n$ - 3) + ($n$ - $n$ - 2) + ($n$ + $n$ - 1) + ($n$ - 1) \\

Simplifying the above formula, we get
\begin{equation}
T_m = n (n + 1) / 2
\end{equation}
which is of O(n$^2$).

In our modified algorithm, considering worst case and assuming lowest process start election, then:
\end{enumerate}
\begin{enumerate}
\item Total number of election message sent to set ($S$) of $n$ processes (\{$P_1$, $P_2$, $P_3$, \dots , $P_n$\}) are ($n$ - 1).
\item Total response message received is ($n$ - 1).
\item Informing to coordinator and coordinator to check with past coordinator involve two messages, and
\item Finally informing to every process by sending coordinator message is again ($n$ - 1) message.
\end{enumerate}
The number of message passing between processes for performing election is obtained from the following formula:\\
$T_m$ = ($n$ -1) + ($n$ - 1) + 1 + 1 + ($n$ - 1), or
\begin{equation}
T_m = 3n -1~ or~ 3n
\end{equation}
 which is of  O($n$).

\section{\uppercase{Conclusions}}
In this paper, we discussed the drawbacks of Bully algorithm and then we presented an optimized method for the Bully algorithm called modified bully algorithm.  Modified Bully algorithm shows improved performance than the Bully algorithm. The additional advantages of modified Bully algorithm are that this algorithm is a very simple, having fail-safe mechanism, no parallel election, and reduced number of messages.
\vskip 2mm
Our analytical simulation shows that our algorithm is more efficient rather than the Bully algorithm, in both number of message passing and the number of stages, and when only one process runs the algorithm message passing complexity decreased from O($n^2$) to O($n$). In this analysis we consider the worst case in modified algorithm. Result of this analysis clearly shows that modified algorithm is better than bully algorithm with fewer message passing and the fewer stages.
\small
\balance

\noindent{\includegraphics[width=1in,height=1.7in,clip,keepaspectratio]{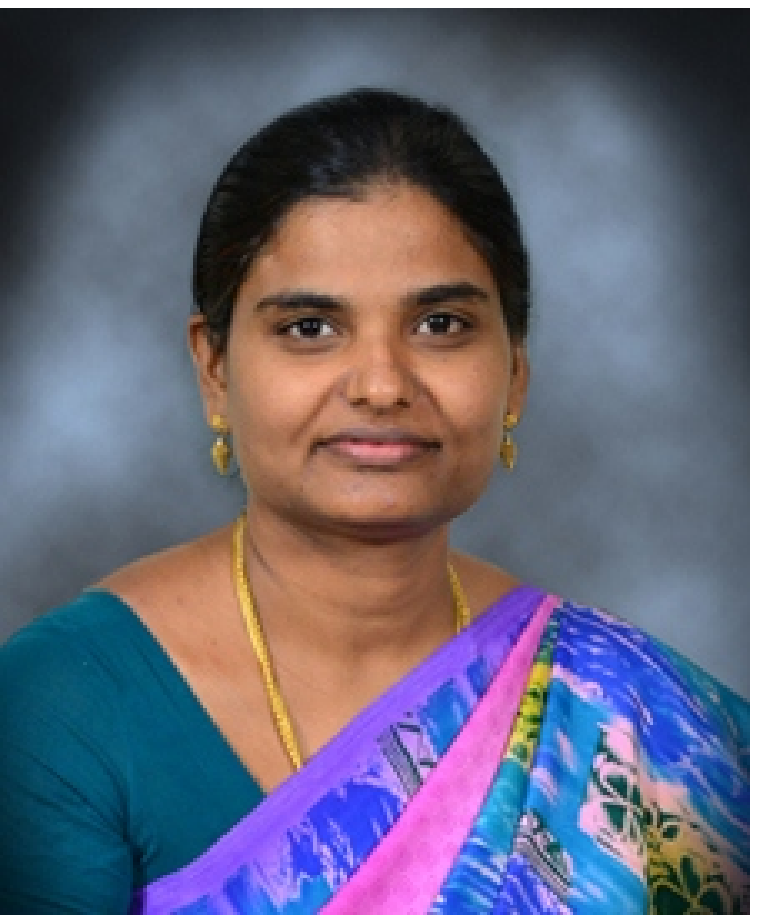}}
\begin{minipage}[b][1in][c]{1.8in}
{\centering{\bf {P  Beaulah  Soundarabai}}  is presently an Assistant Professor in the Department of Computer Science in Christ University, Bangalore. She obtained her Bachelors, Masters Degrees in Computer Applications from Madurai Kamarai University, }\\
\end{minipage} Madurai. She also has completed M. Phil. in Computer Science. Presently she is pursuing Ph.D in
the field of Distributed Systems in Christ University. \\\\
\noindent{\includegraphics[width=1in,height=1.7in,clip,keepaspectratio]{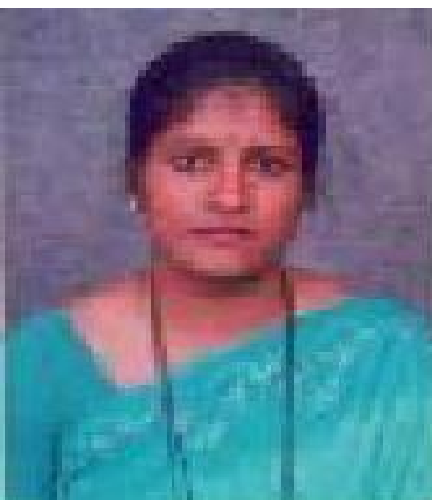}}
\begin{minipage}[b][1in][c]{1.8in}
{\centering{\bf {Thriveni J}} has completed Bachelor of Engineering, Masters of Engineering and Doctoral Degree in Computer Science and Engineering. She has 4 years of industrial experience and 16 years of teaching experience. Currently she is an Associate}\\
\end{minipage}
Professor in the Department of Computer Science and Engineering, University Visvesvaraya College of Engineering, Bangalore. Her research interests include Networks, Data Mining and Biometrics. \\\\

\noindent{\includegraphics[width=1in,height=1.8in,clip,keepaspectratio]{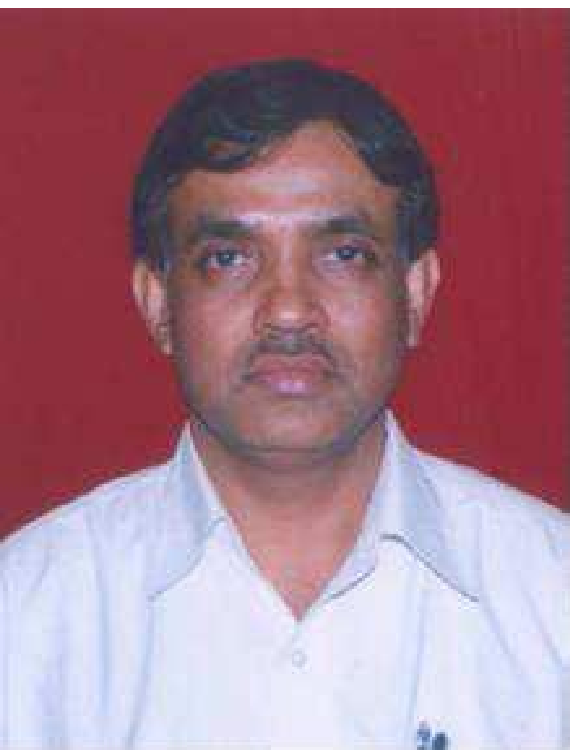}}
\begin{minipage}[b][1in][c]{1.8in}
{\centering{\bf {Venugopal K R}} is currently the Principal, University Visvesvaraya College of Engineering, Bangalore University, Bangalore. He obtained his Bachelor of Engineering from University Visvesvaraya College of Engineering. He received his Masters degree in Computer Science and} \\ \\
\end{minipage}
Automation from Indian Institute of Science Bangalore. He was awarded Ph.D in Economics from Bangalore University and Ph.D in Computer Science from Indian Institute of Technology, Madras. He has a distinguished academic career and has degrees in Electronics, Economics, Law, Business Finance, Public Relations, Communications, Industrial Relations, Computer Science and Journalism. He has authored and edited 39 books on Computer Science and Economics, which include Petrodollar and the World Economy, C Aptitude, Mastering C, Microprocessor Programming, Mastering C++ and Digital Circuits and Systems $etc.$. During his three decades of service at UVCE he has over 400 research papers to his credit. His research interests include Computer Networks, Wireless Sensor Networks, Parallel and Distributed Systems, Digital Signal Processing and Data Mining.\\

\noindent{\includegraphics[width=1in,height=1.8in,clip,keepaspectratio]{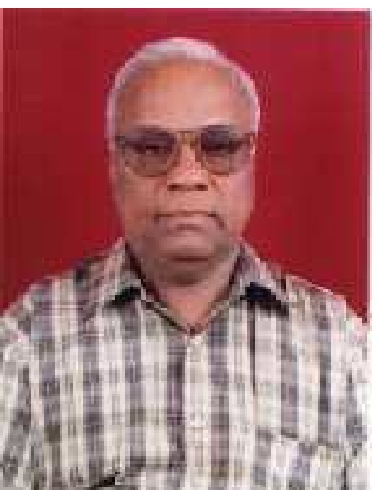}}
\begin{minipage}[b][1in][c]{1.8in}
{\centering{\bf {L M Patnaik}} is currently Honorary Professor, Indian Institute of Science, Bangalore, India. He was a Vice Chancellor, Defense Institute of Advanced Technology, Pune, India and was a Professor since 1986 with the Department of Computer Science and Automation, Indian} \\ \\
\end{minipage}
Institute of Science, Bangalore. During the past 35 years of his service at the Institute he has over 700 research publications in refereed International Journals and refereed International Conference Proceedings. He is a Fellow of all the four leading Science and Engineering Academies in India;  Fellow of the IEEE and the Academy of Science for the Developing World. He has received twenty national and international awards; notable among them is the IEEE Technical Achievement Award for his significant  contributions to High Performance Computing and Soft Computing. His areas of research interest have been Parallel and Distributed Computing, Mobile Computing, CAD for VLSI circuits, Soft Computing and Computational Neuroscience. \\\\\\
\balance

\begin{thebibliography}{00}

\bibitem{1}
{M R Effat Parvar}. {Improved Algorithms for Leader Election in Distributed Systems},  \emph{in 2nd International Conference on Computer Engineering and Technology (ICCET)}, 2010.

\bibitem{2}
{Sandipan Basu}. {An Efficient Approach of Election Algorithm in Distributed Systems}, \emph{Indian Journal of Computer Science and Engineering (IJCSE)}, 2:16--21, 2011.

\bibitem{3}
{Muhammad Mahbubur Rahman and Afroza Nahar}. {Modified Bully Algorithm using Election Commission}, \emph{MASAUM Journal of Computing(MJC)}, 1(3):439--446, October 2009.

\bibitem{4}
{Chang-Young Kim and Sung-Hoon Bauk}. {The Election Protocol for Reconfigurable Distributed Systems}, in \emph{{International Conference on Wireless Networks (ICWN)}},  pages 295--301, 2006.

\bibitem{5}
{M S Kordafshari, M Gholipour, M Jahanshahi and A T Haghighat}. {Modified Bully Election Algorithm in Distributed System}, in \emph{{WSEAS Conferences}}, 2005.

\bibitem{6}
{M Sepehri and M Goodarzi}. {Leader   Election   Algorithm   using   Heap Structure},  \emph{{Proceedings of the 12th WSEAS  international conference on Computers (ICCOMP’08)}},  2008.

\bibitem{7}
{Sung-Hoon Park}. {TA Stable Election Protocol based on an Unreliable Failure Detector in Distributed Systems}, \emph{{Proceedings of IEEE Eighth International Conference on Information Technology: New Generations}},  pages 976--984, 2011.

\bibitem{8}
{M Zargarnataj}. {New Election Algorithm based on Assistant in Distributed Systems}, \emph{{IEEE International Conference on Computer Systems and  Applications}},  pages 324--331, 2007.

\bibitem{9}
{H Garcia-Molina}. {Elections in Distributed Computing System}, \emph{{IEEE Transaction on Computers}},  310:48-59, 1982.

\bibitem{10}
{Chang Ben Ari}. {Principles of Concurrent and Distributed Programming}, \emph{{Pearson Education}},  2nd edition, 2006.

\bibitem{11}
{Andrew S Tanenbaum}. {Distributed Systems Principles and Paradigms}, \emph{{Beijing: Tsinghua University Press}}, 2008.

\bibitem{12}
{G Le Lan}. {Distributed System Towards a Formal Approach}, \emph{{Information Processing, The Netherlands: North-Holland}}, pages 155--160, 1977.

\bibitem{13}
{ Sung-Hoon-Park}. {A Probablistically Correct Election Protocol in Asynchronous Distributed System}, \emph{{APPT, LNCS}}, 310, pages 177--183, 2003.\\\\
\small
\balance
\end{thebibliography}
\end{document}